\documentclass[conference]{IEEEtran}
\usepackage{cite}
\usepackage{amsmath,amssymb,amsfonts}
\usepackage{algorithmic}
\usepackage{graphicx}
\usepackage{textcomp}
\usepackage[T1]{fontenc}

\ifCLASSINFOpdf
\else
\fi
\begin{document}
\title{Non-RLL DC-Balance based on a Pre-scrambled Polar Encoder for Beacon-based Visible Light Communication Systems}

\author{\IEEEauthorblockN{Duc-Phuc NGUYEN, Dinh-Dung LE, Thi-Hong TRAN, Yasuhiko NAKASHIMA}
Graduate School of Science and Technology\\
Nara Institute of Science and Technology\\
630-0192, Takayama-cho 8916-5 Ikoma-shi, Nara-ken, Japan\\
Email: (nguyen.phuc.ni6, le.dung.ku9, hong, nakashim)@is.naist.jp}


%Telephone: (800) 555--1212, Fax: (888) 555--1212}


\maketitle

\begin{abstract}
Current flicker mitigation (or DC-balance) solutions based on run-length limited (RLL) decoding algorithms are high in complexity, suffer from reduced code rates, or are limited in application to hard-decoding forward error correction (FEC) decoders. Fortunately, non-RLL DC-balance solutions can overcome the drawbacks of RLL-based algorithms, but they meet some difficulties in system latency, low code rate or inferior error-correction performance. Recently, non-RLL flicker mitigation solution based on Polar code has proved to be a most optimal approach due to its natural equal probabilities of short runs of 1's and 0's with high error-correction performance. However, we found that this solution can only maintain DC balance only when the data frame length is sufficiently long. Therefore, these solutions are not suitable for using in beacon-based visible light communication (VLC) systems, which usually transmit ID information in small-size data frames. In this paper, we introduce a flicker mitigation solution designed for beacon-based VLC systems that combines a simple pre-scrambler with a (256;158) non-systematic polar encoder.
\end{abstract}

% \fi
\IEEEpeerreviewmaketitle

\section{Introduction}
\IEEEPARstart{V}{LC} simultaneously provides both illumination and communication services. Current VLC systems use On-Off Keying (OOK) modulation to convey digital information via light-emitting diode (LED) signals. The brightness and stability of the light are affected by the distribution of the 1\char`\'s and 0\char`\'s in the data frames \cite{IEEEhowto:kopk18}. Therefore, DC-balance (or flicker mitigation) solutions to maintain approximately equal numbers of zero and one bits in the data frames of VLC systems are an essential concern. 

Tab. I summarizes proposals related to FEC and flicker mitigation for VLC. The conventional solution is defined in the IEEE 802.15.7 VLC standard \cite{IEEEhowto:kopk1} and employs Reed-Solomon (RS) codes, Convolutional Codes (CC) and hard-decoding RLL (hard-RLL) algorithms. However, the applicability of hard-RLL methods \cite{IEEEhowto:kopk1,IEEEhowto:kopk2,IEEEhowto:kopk3} is limited to hard-decoding FEC; consequently, the error-correction performance of the entire system is restricted. Recently, soft-decoding RLL (soft-RLL) solutions have been proposed in \cite{IEEEhowto:kopk4,IEEEhowto:kopk5,IEEEhowto:kopk6,IEEEhowto:kopk16}. These techniques permit soft-decoding FEC algorithms to be applied to improve the bit-error-rate (BER) performance of VLC systems, but they also require high computational effort, with many additions and multiplications \cite{IEEEhowto:kopk15}. 

Zunaira \emph{et al.} have proposed replacing the classic RLL codes with a recursive Unity-Rate Code (URC) as the inner code and a 17-subcode IRregular Convolutional Code (IRCC) as the outer code \cite{IEEEhowto:kopk7}. Although this method can achieve a dimming level of approximately 50\% with good BER performance, the system latency is increased with the iterative-decoding IRCC-URC scheme. In addition, the long codeword length, which ranges from 1000 to 5000 bits, reduces the compatibility of this proposal with beacon-based VLC systems \cite{IEEEhowto:kopk8}. As an alternative approach, Kim \emph{et al.} have proposed two coding schemes based on modified Reed-Muller (RM) codes \cite{IEEEhowto:kopk9}. Although this method can guarantee DC balance at exactly 50\%, it has the inherent drawbacks of a low code rate and inferior error-correction performance compared with turbo codes, low-density parity-check (LDPC) codes or polar codes. In addition, Lee and Kwon have proposed the use of puncturing and pseudo-noise sequence scrambling with compensation symbols (CSs) \cite{IEEEhowto:kopk10}. This proposal can achieve very good BER performance; however, puncturing with CSs will lead to redundant bits in the messages, thereby reducing the transmission efficiency. Another coding scheme based on fountain code, with greatly improved transmission efficiency, was mentioned in \cite{IEEEhowto:kopk11}. However, this scheme requires feedback information and thus is not suitable for broadcasting scenarios in beacon-based VLC systems. 

Finally, Fang \emph{et al.} have recently proposed a non-RLL polar-code-based solution for dimmable VLC \cite{IEEEhowto:kopk12}. This solution offers improved transmission efficiency while achieving a high coding gain compared with RS and LDPC codes. We have found that this solution can overcome the drawbacks of the works mentioned above; it offers non-iterative decoding, a adaptive code rate, and high BER performance without requiring feedback information. The greatest obstacle to this proposal is simply equal probabilities of short runs of 1\char`\'s and 0\char`\'s can only be achieved with a long codeword length (chosen to be \textit{N}=2048 in the cited paper). However, in low-data-rate VLC systems, such as beacon-based ones, long data frames must be avoided. In a beacon-based VLC system, unique ID information is transmitted for purposes such as identifying objects and locations [8]. The 158-bit data frame structure of the proposed beacon-based VLC system is illustrated in Fig. 1. It is evident that a solution \cite{IEEEhowto:kopk12}  based on a polar encoder alone is not applicable in such beacon-based VLC systems because DC balance is not guaranteed for short data frames. 
In this paper, we propose a fast-convergence non-RLL DC-balance scheme based on a pre-scrambled non-systematic polar encoder (NSPE). Consequently, the proposed method can guarantee DC balance in the range of (41.25\%, 63.75\%) for a codeword length of \textit{N}=256, which is 8 times shorter than the codeword length of \textit{N}=2048 required for the polar-code-based solution proposed in \cite{IEEEhowto:kopk12}.

\section{Proposed method}
% non-RLL DC-balance technique for beacon-based VLC systems seems to be a hard problem due to %the %all size of beacon-based data frames. Therefore, we try to solve the remaining problems of %work [9] %y proposing concatenating a pre-scrambler with non-systematic polar encoder.
In a digital transmission system, a data scrambler plays an important role because it causes energy to be spread more uniformly. At the transmitter, a pseudorandom cipher sequence is modulo-2 added to the data sequence to produce a scrambled data sequence. 

Describe the generating polynomial P(x) as:
\begin{equation}\label{eq:polynomial}
P(x)=\displaystyle\sum_{q=0}^{N} c_q.x^q
\end{equation}
where $c_0$ = 1 and is equal 0 or 1 for other indexes.

We have found that the output bit probability distributions of pre-scramblers with different generating polynomials seem to differ slightly. Therefore, we propose the simple generating polynomial presented in (\ref{eq:scrambler}) to reduce the number of shift registers required in the pre-scrambler. 
\begin{equation}\label{eq:scrambler}
P(x)=x^4 + x^3 + 1
\end{equation}

Meanwhile, polar codes can be classified into two types: non-systematic and systematic codes. Typically, a polar code is specified by a triple consisting of three parameters: \textit{(N, K, I)}, where \textit{N} is the code length, \textit{K} is the message length, and \textit{I} is the set of information bit indices. Let \textit{d} be a vector of \textit{N} bits, including information bits. The generator matrix is defined as \(G=(F^{\otimes n})_I\). Then, given a scrambled message \textit{u} of \textit{K} bits in length, a codeword \textit{x} is generated as given in (\ref{eq:polarencode}). 
\begin{equation}\label{eq:polarencode}
x = u.G = d.F^{\otimes n}
\end{equation}
Systematic polar codes were introduced to achieve better error-correction performance compared with non-systematic polar codes \cite{IEEEhowto:kopk13}. However, we have found that because the information bits transparently appear as part of the codewords, the output probability distribution of a systematic polar encoder is not well centralized. By contrast, an NSPE is formed of many layers of XOR gates, with a complexity of \(\frac{N}{2}log_2N\) XORs. The output bit probability distribution of an NSPE naturally becomes centralized at approximately 50\% 1\char`\'s and 50\% 0\char`\'s as the codeword length increases \cite{IEEEhowto:kopk12}. 
We select an NSPE as the basis for our main FEC scheme for beacon-based VLC systems for several reasons:
\begin{enumerate}
\item The encoder's output bit probability distribution is naturally centralized at 50\% 1\char`\'s and 50\% 0\char`\'s.
\item Unusual code rates are supported. Specifically, a (256;158) polar code, which has a code rate of 0.617, is suitable for a beacon-based VLC frame size of \textit{K}=158.
\item High error-correction performance can be achieved with low hardware complexity \cite{IEEEhowto:kopk14,IEEEhowto:kopk17}.
\item The inherently short run lengths of a polar encoder can mitigate lighting flicker \cite{IEEEhowto:kopk12}.
\item An NSPE is less affected by the input distribution of 1\char`\'s and 0\char`\'s than a systematic polar encoder.
\end{enumerate}
%However, beacon-based frame size is relatively small (158-bit) and current non-RLL solutions %hich often require long data frames, seem to be not suitable for such beacon-based VLC systems. 
A pre-scrambler can help to ensure the fast convergence of the output probability distribution of an inner (256;158) NSPE. As a result, DC balance in a beacon-based VLC system can be guaranteed by the proposed system depicted in Fig. 1. Since the proposed method is coding-based flicker mitigation, OOK modulation is considered because of its simplicity. Also, at the receiver, we are trying to implement a filter in which log-likelihood ratio (LLR) values are calculated from probability information of received bits.

Fig.2 shows the hardward block diagram of the proposed pre-scrambled Polar encoder. A Polar encoder which has codeword length N=16 is selected for easier illustration of the proposed architecture. Specifically, sequential serial architecture is used for the pre-scrambler while a recursive architecture based on convolutional logic is used to implement the Polar encoder (256,158).

\begin{figure}[!t]
\centerline{\includegraphics[width=\columnwidth]{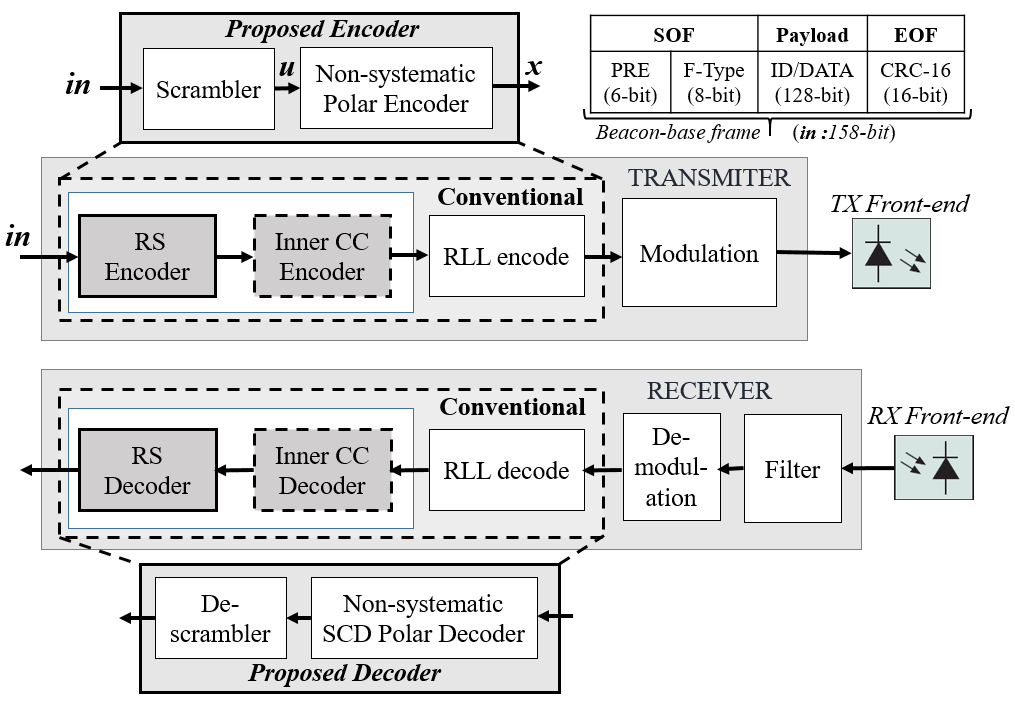}}
\caption{The proposed beacon-based VLC system}
\label{fig1}
\end{figure}
\begin{figure}[!t]
\centerline{\includegraphics[width=\columnwidth]{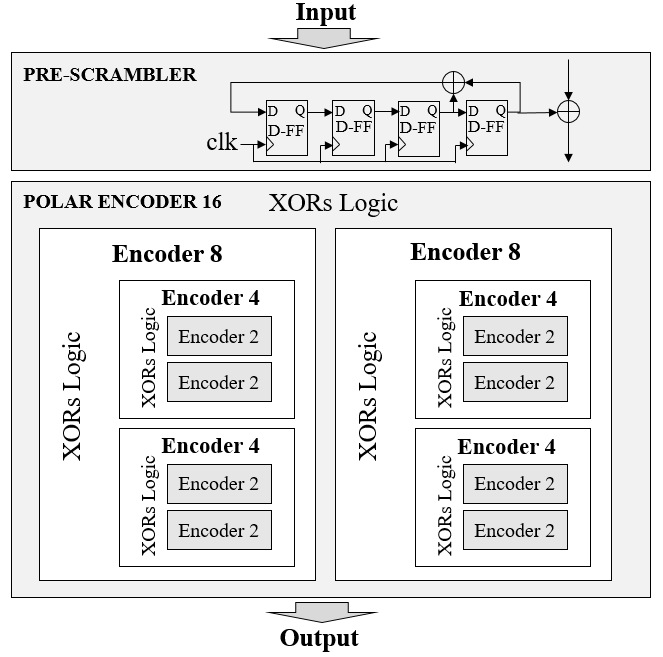}}
\caption{The proposed pre-scrambled non-systematic polar encoder }
\label{fig2}
\end{figure}

\begin{table}[!t]
\renewcommand{\arraystretch}{1.3}
\caption{Overview of FEC and flicker mitigation algorithms for VLC}
\label{table_1}
\centering
\begin{tabular}{c||c}
\hline
\bfseries FEC Solution & \bfseries Flicker Mitigation\\
\hline\hline
RS, CC \cite{IEEEhowto:kopk1} & Hard-RLL\\\hline
Multi-RS hard-decoding \cite{IEEEhowto:kopk2}&Hard-RLL\\\hline
LDPC \cite{IEEEhowto:kopk3}&Hard-RLL \\\hline
RS soft-decoding \cite{IEEEhowto:kopk4,IEEEhowto:kopk5}&Soft-RLL \\\hline
Polar code \cite{IEEEhowto:kopk6,IEEEhowto:kopk16}&Soft-RLL \\\hline
Irregular CC \cite{IEEEhowto:kopk7}& Unity-Rate Code\\\hline
Reed-Muller \cite{IEEEhowto:kopk9}& Modified original code\\\hline
Turbo code \cite{IEEEhowto:kopk10}& Puncture + Scrambling\\\hline
Fountain code \cite{IEEEhowto:kopk11}& Scrambling\\\hline
Polar code ($N$=2048) \cite{IEEEhowto:kopk12}& Flicker-free\\\hline
Pre-scrambled polar code\\ 
($K$=158, $N$=256 beacon frame size),\\
proposed method& Flicker-free\\\hline
\hline
\end{tabular}
\end{table}

\section{Results}
The maximum flickering time period (MFTP < 5 ms) is defined as the maximum time period over which the intensity of light can change without being perceived by the human eye. At the lowest optical clock rate permitted by the IEEE VLC standard, namely, 200 kHz, 256 encoded bits will be transmitted in 0.79 ms, which is far shorter than the MFTP. Thus, we can assume that no potential flicker will be induced because even the maximum run length can be accepted.  
For the previously proposed non-RLL solution based only on a polar encoder \cite{IEEEhowto:kopk12}, the authors demonstrated the fluctuation of the code weight distribution around the 50\% dimming level. Specifically, in the case of a polar encoder with 2048-bit codewords, the percentage of one bits was reported to fluctuate in the range of (42.1875\%, 57.8125\%).
%%%Senior Editor - Please ensure that the intended meaning has been maintained
%%%in the above edit.
However, we have found that this fluctuation range can only be achieved when the proportions of 1\char`\'s and 0\char`\'s in the input data (before the FEC encoder) are both equal to approximately 50\%. Unfortunately, the bit ratio of the input data is unknown beforehand because of the randomness of these data, and this input bit ratio greatly affects the output bit ratio of the FEC encoder. In this paper, we evaluate our proposed method using a worst-case input bit ratio corresponding to 10\% zero bits and 90\% one bits. A simulation was performed using 10,000 158-bit data frames. If the minimum and maximum bit ratios are included, the real fluctuation range of the output bit probability distribution of a polar encoder with 2048-bit codewords is (41.25\%, 61.25\%). 
From the experimental results presented in Fig. 3, we can also see that an NSPE shows a more centralized bit distribution compared with that of a systematic encoder regardless of whether pre-scrambling is applied. Especially when a pre-scrambler is not used, the probability distribution of the systematic polar encoder tends toward 85\% one bits
%%%Senior Editor - Please ensure that the intended meaning has been maintained
%%%in the above edit.
because it is greatly affected by the input bit probability distribution. Therefore, we adopt an NSPE as the basis of the main FEC scheme in our proposal. Fig. 4 shows the impact of a pre-scrambler on the output bit probability distribution of the NSPE. Notably, DC balance is not guaranteed in the case of a (256;158) polar code if a pre-scrambler is not applied because the encoder's output bit probability distribution is spread over a large range of percentages (32.5\%, 85\%). However, when a pre-scrambler is used, the output fluctuation range of the pre-scrambled (256;158) polar encoder is (41.25\%, 63.75\%), whereas the fluctuation ranges of polar encoders with codeword lengths of 2048 and 1024 are (41.25\%, 61.25\%) and (38.75\%, 67.5\%), respectively. Thus, pre-scrambling causes the output bit probability distribution of a (256;158) NSPE to be approximately equal to those of (1024;512) and (2048;1024) encoders. In other words, considering the frame size of beacon-based systems, a pre-scrambler is necessary to ensure faster convergence to a centralized bit probability distribution.

The experimental results presented in Fig. 3 and 4 demonstrate that a pre-scrambler combined with an NSPE is useful for ensuring faster convergence to a centralized output bit probability distribution. Accordingly, DC balance can be guaranteed in beacon-based VLC systems with a short data frame length of 158 bits. Compared with the non-RLL DC-balance solution based only on a polar encoder with 2048-bit codewords presented in \cite{IEEEhowto:kopk12}, the proposed method can achieve the same output bit probability distribution with a codeword length that is shorter by a factor of 8. Moreover, in our previous work \cite{IEEEhowto:kopk14,IEEEhowto:kopk17}, we have demonstrated that a successive-cancellation polar decoder is low in complexity and has better BER performance compared with the conventional RS decoding solutions in the VLC PHY layer defined in the IEEE standard \cite{IEEEhowto:kopk1}. Specifically, fig.5 shows the BER performance of the soft-decoding of pre-scrambled non-systematic Polar code (256;158). We can found that the BER performance of Polar code in code-rate = 0.62 outperforms RS code at code-rates (15/11), (15/7) and (15/3) mentioned in \cite{IEEEhowto:kopk1,IEEEhowto:kopk4} with coding gains at BER = $10^{-5}$ are 3.4, 4.2 and 6.4 respectively.    

\section{Conclusion}
We have proposed a non-RLL DC-balance solution consisting of a pre-scrambler based on a simple generating polynomial combined with an NSPE. The proposed method has a centralized bit probability distribution, with approximately equal numbers of zero and one bits; therefore, DC balance can be maintained even with the short data frames used in beacon-based VLC systems. Moreover, the non-RLL nature of the proposal enables soft decoding of polar codes, which can be applied in VLC receivers to enhance the error-correction performance. 
\begin{figure}[!t]
\centerline{\includegraphics[width=\columnwidth]{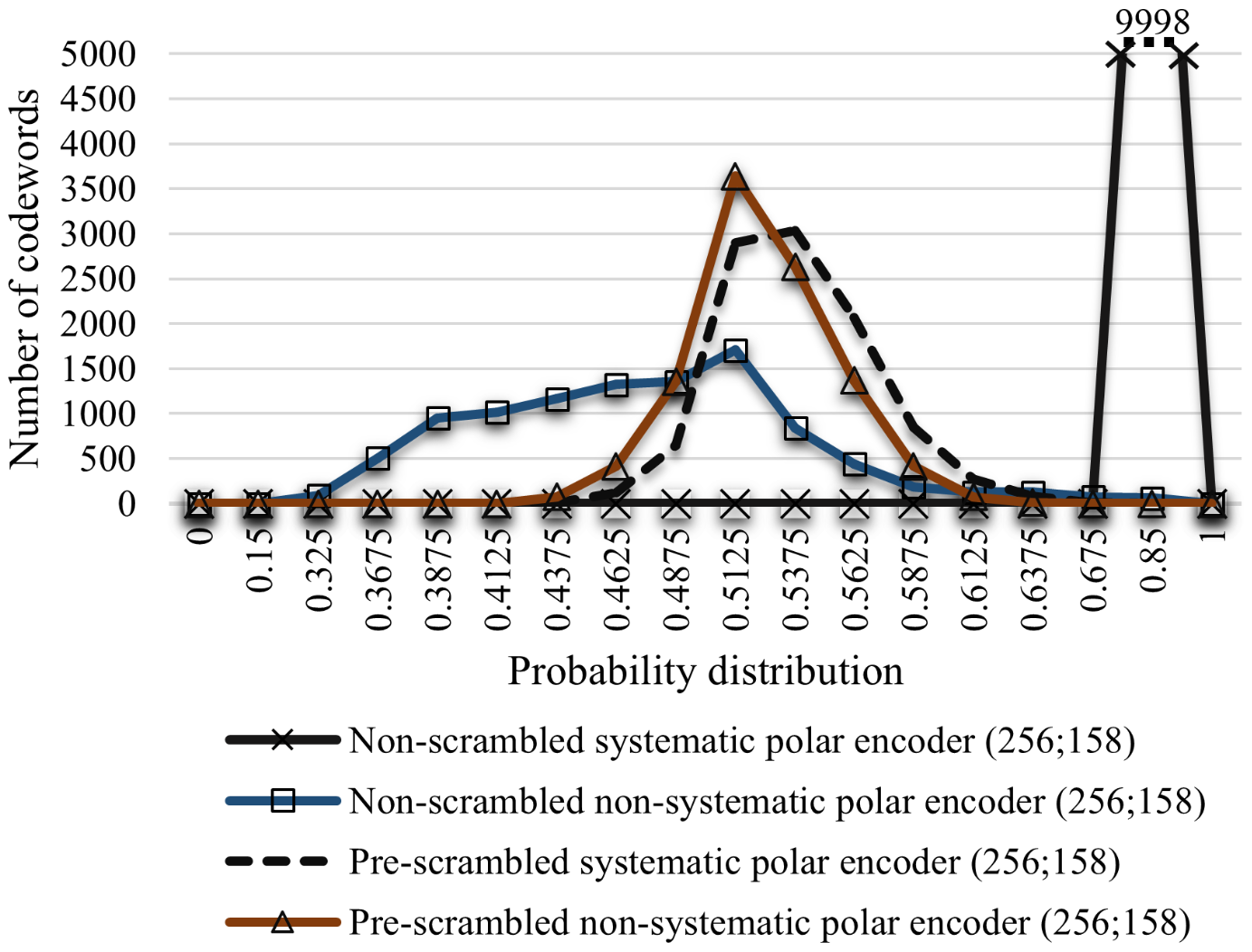}}
\caption{Output bit probability distributions of non-systematic and systematic polar encoders }
\label{fig3}
\end{figure}

\begin{figure}[!t]
\centerline{\includegraphics[width=\columnwidth]{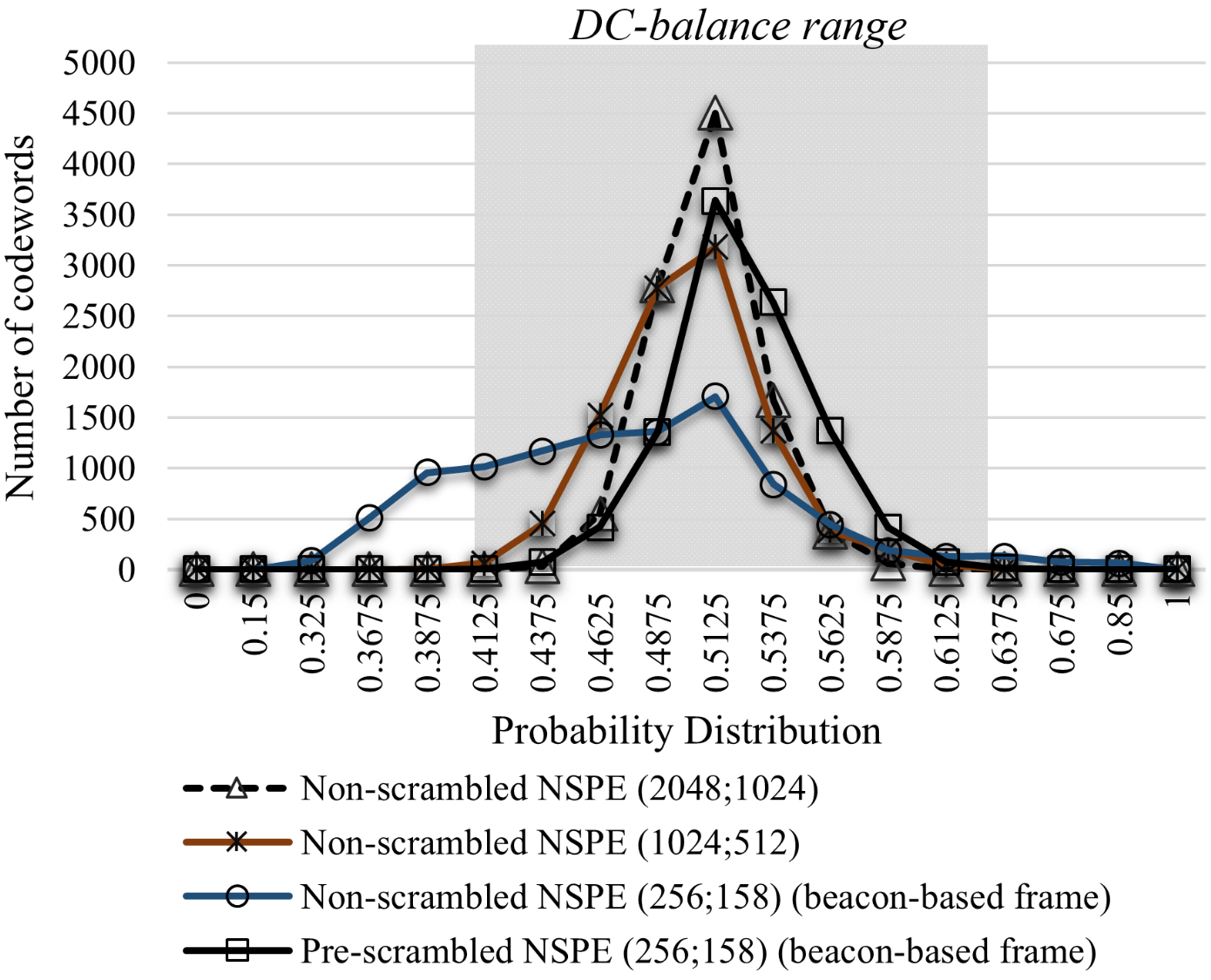}}
\caption{Output bit probability distributions of pre-scrambled and non-scrambled NSPEs with long and short codeword lengths}
\label{fig4}
\end{figure}

\begin{figure}[!t]
\centerline{\includegraphics[width=\columnwidth]{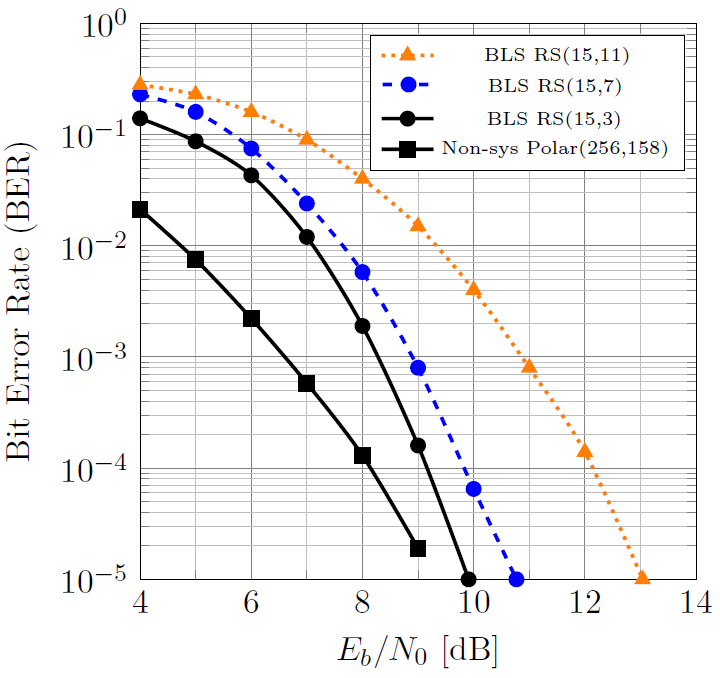}}
\caption{Bit-error-rate performance of soft-decoding of pre-scrambled non-systematic Polar code (256;158)}
\label{fig5}
\end{figure}


\section*{Acknowledgment}
This work  was supported by both Creative and International Competitiveness Project (CICP 2017) and JSPS KAKENHI Grant Number JP16K18105.



%\bibliographystyle{IEEEtran}

\end{document}